\documentclass{emulateapj}

\usepackage{natbib}
\usepackage{graphicx}
\usepackage{dcolumn}
\usepackage{bm}

\shorttitle{Detecting the Milky Way's Dark Disk}
\shortauthors{Bruch et al.}

\begin{document}

\def \apj  {ApJ}
\def \apjl  {ApJL}
\def \aj  {AJ}
\def\mnras {MNRAS}
\def \etal {{\it et~al.}}
\def \chisq  {\ifmmode  \chi^2   \else  $\chi^2$  \fi}  
\def \chisqr {\ifmmode \chi^2_{\rm r} \else $\chi^2_{\rm r}$ \fi}
\def \spose#1{\hbox  to 0pt{#1\hss}}  
\def \lta{\mathrel{\spose{\lower 3pt\hbox{$\sim$}}\raise  2.0pt\hbox{$<$}}}
\def \gta{\mathrel{\spose{\lower  3pt\hbox{$\sim$}}\raise 2.0pt\hbox{$>$}}}
\def \ion#1#2{#1{\footnotesize{#2}}\relax} 
\def \ha  {\ifmmode H\alpha \else H$\alpha $ \fi} 
\def \hi {\ion{H}{I}} 
\def \hii {\ion{H}{II}} 
\def \oii {[\ion{O}{II}]} 

\def \kmsmpc {\>{\rm km}\,{\rm s}^{-1}\,{\rm Mpc}^{-1}}
\def \kms {\ifmmode  \,\rm km\,s^{-1} \else $\,\rm km\,s^{-1}  $ \fi }
\def \kpc {\ifmmode  {\rm kpc}  \else ${\rm  kpc}$ \fi  }  
\def \Msun {\ifmmode M_{\odot} \else $M_{\odot}$ \fi} 
\def \Mearth {\ifmmode M_{\oplus} \else $M_{\oplus}$ \fi} 
\def \hMsun {\ifmmode h^{-1}\,\rm M_{\odot} \else $h^{-1}\,\rm M_{\odot}$ \fi}
\def \hhMsun {\ifmmode h^{-2}\,\rm M_{\odot}\else $h^{-2}\,\rm M_{\odot}$ \fi}
\def \Lsun {\ifmmode L_{\odot} \else $L_{\odot}$ \fi} 
\def \hhLsun {\ifmmode h^{-2}\,\rm L_{\odot} \else $h^{-2}\,\rm L_{\odot}$ \fi}

\def \LCDM {\ifmmode \Lambda{\rm CDM} \else $\Lambda{\rm CDM}$ \fi}
\def \sig8 {\ifmmode \sigma_8 \else $\sigma_8$ \fi} 
\def \OmegaM {\ifmmode \Omega_{\rm M} \else $\Omega_{\rm M}$ \fi} 
\def \OmegaL {\ifmmode \Omega_{\rm \Lambda} \else $\Omega_{\rm \Lambda}$\fi} 
\def \Deltavir {\ifmmode \Delta_{\rm vir} \else $\Delta_{\rm vir}$ \fi}
\def\lya{Lyman-$\alpha$}
\def\hang{\par\hangindent=\parindent\noindent}

\def \LCDM {\ifmmode \Lambda{\rm CDM} \else $\Lambda{\rm CDM}$\fi}
\def \lCDM {\ifmmode \Lambda{\rm CDM} \else $\Lambda{\rm CDM}$ \fi}
\def \lcdm {\ifmmode \Lambda{\rm CDM} \else $\Lambda{\rm CDM}$\fi}

\def \rs {\ifmmode r_{\rm s} \else $r_{\rm s}$ \fi} 
\def \rrm2 {\ifmmode r_{-2} \else $r_{-2}$ \fi} 
\def \ccm2 {\ifmmode c_{-2} \else$c_{-2}$ \fi} 
\def \cvir {\ifmmode c_{\rm vir} \else $c_{\rm vir}$ \fi} 
\def \cbar {\ifmmode \overline{c} \else $\overline{c}$ \fi}

\def \R200 {\ifmmode R_{200} \else $R_{200}$ \fi} 
\def \Rvir {\ifmmode R_{\rm vir} \else $R_{\rm vir}$ \fi}

\def \v200 {\ifmmode V_{200} \else $V_{200}$ \fi} 
\def \Vvir {\ifmmode V_{\rm  vir} \else  $V_{\rm vir}$  \fi} 
\def  \Vhalo  {\ifmmode V_{\rm halo} \else $V_{\rm halo}$ \fi}

\def \M200 {\ifmmode M_{200} \else $M_{200}$ \fi} 
\def \Mvir {\ifmmode M_{\rm  vir} \else $M_{\rm  vir}$ \fi}  
\def \Mshell  {\ifmmode M_{\rm shell} \else $M_{\rm shell}$ \fi}

\def \Nvir {\ifmmode N_{\rm  vir} \else $N_{\rm  vir}$ \fi}  

\def \Jvir {\ifmmode J_{\rm vir} \else $J_{\rm vir}$ \fi} 
\def \Jshell {\ifmmode J_{\rm shell} \else $J_{\rm shell}$ \fi}

\def \Evir {\ifmmode E_{\rm vir} \else $E_{\rm vir}$ \fi} 

\def \lam {\ifmmode \lambda  \else $\lambda$ \fi} 
\def \lamp {\ifmmode \lambda^{\prime} \else $\lambda^{\prime}$  \fi} 
\def \lampc {\ifmmode \lambda^{\prime}_{\rm c} \else
  $\lambda^{\prime}_{\rm c}$  \fi} 
\def \lambar {\ifmmode \bar{\lambda}  \else  $\bar{\lambda}$  \fi}  
\def  \lampbar  {\ifmmode \bar{\lambda^{\prime}} \else
  $\bar{\lambda^{\prime}}$\fi} 
\def \siglam {\ifmmode \sigma_{\lambda} \else $\sigma_{\lambda}$ \fi} 
\def \siglamp {\ifmmode                \sigma_{\lambda^{\prime}} \else
$\sigma_{\lambda^{\prime}}$\fi} 
\def  \sigl {\sigma_{\ln\lambda}} 
\def \sigc {\sigma_{\ln\cvir}}

\def \Rd {\ifmmode R_{\rm d} \else $R_{\rm d}$ \fi} 
\def \Rs {\ifmmode R_{\rm s} \else $R_{\rm s}$ \fi}  
\def \Rd {\ifmmode R_{\rm d} \else $R_{\rm d}$ \fi}  
\def \Rcool  {\ifmmode R_{\rm  cool}  \else $R_{\rm cool}$ \fi} 
\def \RIII {\ifmmode  3.2\Rs \else $3.2\Rs$ \fi} 
\def \RII {\ifmmode 2.2\Rs \else $2.2\Rs$  \fi} 
\def \Reff {\ifmmode R_{\rm eff} \else $R_{\rm  eff}$ \fi} 
\def  \rb {\ifmmode r_{\rm b}  \else $r_{\rm b}$ \fi}

\def  \Sigmacrit   {\ifmmode  \Sigma_{\rm  crit}   
\else  $\Sigma_{\rm crit}$\fi} 
\def \Sig0 {\ifmmode \Sigma_{0} \else $\Sigma_{0}$ \fi}

\def \muI {\ifmmode \mu_{0,I} \else $\mu_{0,I}$ \fi}

\def \mgal {\ifmmode m_{\rm gal} \else $m_{\rm gal}$ \fi} 
\def \md {\ifmmode m_{\rm d} \else $m_{\rm d}$ \fi} 
\def \ms {\ifmmode m_{\rm   s}   \else   $m_{\rm   s}$   \fi}   
\def   \mdbar   {\ifmmode {\overline{m}}_{\rm d} \else
  ${\overline{m}}_{\rm d}$ \fi} 
\def \msbar {\ifmmode  \bar{m}_{\rm  s}  \else  $\bar{m}_{\rm s}$
  \fi}  
\def  \Md {\ifmmode M_{\rm d}  \else $M_{\rm d}$ \fi} 
\def  \Ms {\ifmmode M_{\rm s} \else $M_{\rm  s}$ \fi} 
\def \Mb {\ifmmode  M_{\rm b} \else $M_{\rm b}$ \fi} 
\def \Mstar {\ifmmode  M_{\rm star} \else $M_{\rm star}$ \fi}
\def \Mdisk {\ifmmode M_{\rm disk} \else $M_{\rm disk}$ \fi}

\def \Jd {\ifmmode J_{\rm d} \else $J_{\rm d}$ \fi} 
\def \Jb {\ifmmode J_{\rm b} \else $J_{\rm b}$ \fi}  
\def \fb {\ifmmode  f_{\rm b} \else $f_{\rm b}$ \fi}

\def  \jd  {\ifmmode j_{\rm  d}  \else  $j_{\rm  d}$ \fi}  
\def  \jdmd {\ifmmode \frac{j_{\rm  d}}{m_{\rm d}} \else
  $\frac{j_{\rm d}}{m_{\rm d}}$ \fi} 
\def \fj {\ifmmode f_{\rm j} \else $f_{\rm j}$ \fi} 
\def \ft {\ifmmode f_{\rm t}  \else $f_{\rm t}$ \fi} 
\def  \fM {\ifmmode f_{\rm M} \else $f_{\rm M}$ \fi}

\def  \Vd {\ifmmode  V_{\rm  d}  \else $V_{\rm  d}$  \fi} 
\def  \Vcool {\ifmmode V_{\rm cool} \else $V_{\rm cool}$ \fi} 
\def \Vcirc {\ifmmode V_{\rm circ}  \else $V_{\rm circ}$  \fi} 
\def \VIII  {\ifmmode V_{3.2} \else $V_{3.2}$ \fi} 
\def  \VII {\ifmmode V_{2.2} \else $V_{2.2}$ \fi}
\def \Vobs {\ifmmode V_{\rm obs}  \else $V_{\rm obs}$ \fi} 
\def \Vdisk {\ifmmode V_{\rm disk} \else  $V_{\rm disk}$ \fi} 
\def \Vmax {\ifmmode V_{\rm  max} \else  $V_{\rm max}$  \fi} 
\def  \Vmaxobs{\ifmmode V_{\rm max}^{\rm obs}\else  $V_{\rm max}^{\rm
    obs}$\fi}  
\def \Vtot {\ifmmode V_{\rm tot} \else $V_{\rm tot}$  \fi} 
\def \Vrot {\ifmmode V_{\rm rot} \else  $V_{\rm rot}$  \fi} 
\def  \Vflat {\ifmmode  V_{\rm  flat} \else $V_{\rm flat}$ \fi}

\def \Ups {\ifmmode \Upsilon  \else $\Upsilon$ \fi} 
\def \YB {\ifmmode \Upsilon_B \else $\Upsilon_B$ \fi} 
\def \YI {\ifmmode  \Upsilon_I  \else $\Upsilon_I$ \fi} 
\def \DeltaIMF {\ifmmode \Delta_{\rm IMF} \else $\Delta_{\rm IMF}$ \fi}

\def\rhodrat{\rho_\mathrm{d}/\rho_\mathrm{h}}
\def\rhod{\rho_\mathrm{d}}
\def\rhoh{\rho_\mathrm{h}}
\def\GeV{GeV$/$c$^2$}
\def\TeV{TeV$/$c$^2$}
\def\MWIMP{M_\mathrm{WIMP}}

\title{Detecting the Milky Way's Dark Disk}

\author{Tobias Bruch$^*$, Justin Read$^\dagger$, Laura Baudis$^*$ and George Lake$^\dagger$} 
\affiliation{University of Z\"urich, Winterthurerstrasse 190, CH-8057,  Z\"urich, Switzerland}
\altaffiliation[$^*$]{Physics Institute}
\altaffiliation[$^\dagger$]{Institute for Theoretical Physics}

\email{tbruch, justin, lbaudis, lake@physik.uzh.ch}

\begin{abstract}

In the standard model of disk galaxy formation, a dark matter disk forms as massive satellites are preferentially dragged into the disk plane and dissolve. Here, we show the importance of the dark disk for direct dark matter detection. The low velocity of the dark disk with respect to the Earth enhances detection rates at low recoil energy. For weakly interacting massive particle (WIMP) masses $\MWIMP\gtrsim50$\,\GeV, the detection rate increases by up to a factor of 3 in the $5 - 20$\,keV recoil energy range. Comparing this with rates at higher energy is sensitive to $\MWIMP$, providing stronger mass constraints particularly for $\MWIMP\gtrsim100$\,\GeV. The annual modulation signal is significantly boosted and the modulation phase is shifted by $\sim3$ weeks relative to the dark halo. The variation of the observed phase with recoil energy determines $\MWIMP$, once the dark disk properties are fixed by future astronomical surveys. The constraints on the WIMP interaction cross section from current experiments improve by factors of 1.4 to 3.5 when a typical contribution from the dark disk is included.

\end{abstract}


\maketitle

\section{The Milky Way's dark disk}

A mysterious dark matter dominates the matter content of the universe. Although there are no dark matter candidates in the standard model, they are plentiful in extended models. Among these, weakly interacting massive particles (WIMPs; \citep{LW77, Gunn78, Ellis84}), which may arise in supersymmetric extensions of the standard model (SUSY) \citep{Jungman} or in theories that include universal extra dimensions \citep{Cheng02,Hooper07}, stand out as well motivated and detectable.

WIMPs may be detected directly by scattering in a laboratory detector \citep{GW85} or indirectly by their annihilation products from their highest density regions \citep{SS84, Lake90}. In the case of direct detection, we must know the dark matter's phase space structure  to predict rates. In early calculations, the standard halo model (SHM) of the dark matter assumed no rotation and the density distribution was taken to be a spherical isothermal sphere with a core radius of several kpc. More recent modeling includes the cuspier central profiles from \lcdm\ simulations \citep{NFW, Moore98}, producing changes of $O(10\%)$ with respect to the SHM \citep{Kamion98}. Larger boosts have been claimed if dark matter is highly clumped \citep{Green02}, but it is more likely that we live outside a clump, leading to a modest reduction in the local density \citep{Kamion08}.

Simulations containing only dar matter particles have extremely high resolution, 
but they may not be addressing the ``next to leading order" of the model because they do not include the effect of the baryons. Read et al. (2008) recently demonstrated that massive satellites are preferentially dragged into the baryonic disk plane by dynamical friction where they dissolve leaving a thick dark matter disk \citep{Read08, darkdisk1}. The precise properties of the dark disk depend on the stochastic merger history and cosmology. However,  given the expected merger history for a typical Milky Way in \lcdm, they found a dark disk with density in the range $\rhodrat \sim 0.2 - 1$ at the solar neighborhood (where $\rhoh$ is the density of the SHM). The lower bound $\rhodrat = 0.2$ is particularly conservative since it is produced by just one merger of Large Magellanic Cloud mass within 20$^\mathrm{o}$ of the disk plane. In \lcdm\ we expect {\it two} such low inclination mergers per Milky Way (and seven in total at all inclinations).

We may obtain an upper bound on $\rhodrat$ from the kinematics of stars at the solar neighborhood \citep{Oort, Bahcall84}. The latest measurements from Hipparcos give a conservative upper bound of $\rhodrat < 3$, including systematic errors  \citep{Hipp, Statler89}; with $\rhodrat$ being $<2$ likely. If more than half of the thick disk owes to accretion, the likely dark disk density would be near the upper limit. As such we consider $\rhodrat = [0.5,1,2]$ in this paper. The disk density $\rhod$ is an excess over $\rhoh$,  locally increasing the dark matter density.

The kinematic properties of the dark disk can be estimated from the accreted stellar thick disk that forms concurrently. Stellar thick disks are found in the Milky Way and in all well-observed spiral galaxies \citep{Burstein,Gilmore83,Yoachim06}, while at least one  {\it counterrotating} thick disk is strong evidence for an accretion origin \citep{YD05}.  However, thick disks can also form through heating of an underlying thin disk \citep{Kazan}, or even directly from extended gas \citep{Brook}; indeed it is difficult in \lcdm\ to obtain a thick disk massive enough from accretion alone \citep{Read08}. In this paper, we assume -- based on the numerical models of \citep{Read08} -- that the dark disk's kinematics match the Milky Way's stellar thick disk. At the solar neighborhood, this gives a rotation lag $v_{lag}$ of $40-50$\,km $s^{-1}$ with respect to the local circular velocity, and dispersions of $(\sigma_R,\sigma_\phi,\sigma_z) = (63,39,39)$\,km $s^{-1}$ \citep{Read08}. Since the dispersions are nearly isotropic and somewhat uncertain, we model the dark disk with a simple one-dimensional Maxwellian distribution, with a dispersion and lag, $\sigma = v_{lag} = 50$\,km s$^{-1}$ to show its general effect on direct detection. Improving on this assumption will involve untangling heated versus accreted components in the Milky Way stellar thick disk. This should become possible with future astronomical surveys like the Radial Velocity Experiment (RAVE) \citep{Rave06} and Global Astrometric Interferometer for Astrophysics \citep{GAIA} that will provide full 6D phase space and chemical information for hundreds of thousands of individual stars. 

We emphasize that the dark disk must form in any hierarchical model of structure formation; it
is not special to our \lcdm\ simulations \citep{Read08}. However, 
\lcdm\ is specific enough to predict the dark disk density with uncertainties owing only to the stochastic nature of the merging and accretion history.  
The dark disk is very different from dark matter streams \citep{Freese04, Savage06} that have a low filling factor (we are not likely to live in a stream), and are stochastic {\it microstructure}.  By contrast, the dark disk is the expected equilibrium end state of dissolving satellites and the Earth {\it must} be embedded in one (if 
hierarchical formation is correct). Like the near-spherical dark matter halo, the dark disk is {\it macrostructure}.

\begin{figure}
\includegraphics[width=0.47\textwidth]{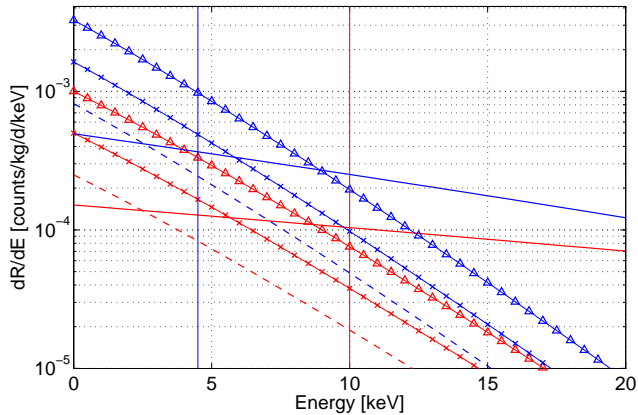}
\caption{Differential recoil rates for Ge (red/gray) and Xe (blue/black) targets, for $\MWIMP=100$ \GeV\ and $\sigma_\mathrm{(WIMP,N)}= 10^{-8}$ pb in the SHM (solid line) and the dark disk. Three different values of $\rhodrat$ (0.5 dashed, 1 {\large{$\times$}} and 2 $\bigtriangleup$) are shown. The vertical lines mark current experiment thresholds: XENON10 (blue/black) using a Xe and CDMS-II (red/gray) using a Ge target.}
\end{figure}

\section{Direct Detection and the Dark Disk}

Direct detection experiments measure nuclear recoil rates above the detector's energy threshold \citep{Baudis}; here we consider Ge and Xe. The detected elastic WIMP-nucleon recoils will range from a few to tens of keV. The expected recoil rate per unit mass,  unit nuclear recoil energy and unit time is \citep{LewinSmith}
\begin{equation}
\frac{dR}{dE} = \frac{\rho \sigma_\mathrm{(WIMP,N)} |F(E)|^2}{2 M_\mathrm{WIMP} \mu^2}  \int_{v>\sqrt{m E/ 2 \mu^2}}^{v_{max}} \frac{f(\textbf{v},t)}{v} d^3v
\end{equation}
\noindent
where $\rho$ is the local dark matter density ($\rho_h = 0.3$ GeV cm$^{-3}$ in the SHM), $\sigma_\mathrm{(WIMP,N)}$ is the WIMP-nucleus scattering cross section, $F(E)$ is the nuclear form factor,  
$\MWIMP$ and $m$ are the masses of the dark matter particle and of the target nucleus, respectively, 
$\mu$ is the reduced mass of the WIMP-nucleus system, $v=|\textbf{v}|$ and $v_{max}$ is the maximal velocity in the earth frame for particles moving at the galactic escape velocity $v_{esc}=544$\,km s$^{-1}$ \citep{Rave07}.  
We only consider the spin-independent (SI) scalar WIMP-nucleus coupling in this paper, since it
dominates the interaction (depending however
on the dark matter particle) for target media with nucleon number $A$ $\gtrsim30$ \citep{Jungman}.
We model the velocity distributions of particles in the dark disk and the SHM with a simple one-dimensional Maxwellian:
\begin{equation}
f(\textbf{v},t) \propto \exp\left(\frac{-(\textbf{v}+\textbf{v}_\oplus(t))^2}{2\sigma^2}\right)
\end{equation}
\noindent
where $\textbf{v}$ is the laboratory velocity of the dark matter particle and the instantaneous streaming velocity
$\textbf{v}_\oplus=\textbf{v}_\mathrm{circ}+\textbf{v}_\odot+\textbf{v}_\mathrm{orb}(t)$. This
streaming velocity is the sum of local circular velocity $\textbf{v}_\mathrm{circ} =(0, 220, 0)$\,km s$^{-1}$, the peculiar motion of the Sun $\textbf{v}_\odot = (10.0, 5.25, 7.17)$\, km s$^{-1}$ \citep{DehnenBinney} with respect to $\textbf{v}_\mathrm{circ}$
and the orbital velocity of the Earth around the Sun $\textbf{v}_\mathrm{orb}(t)$. In the SHM, the halo has no rotation and the dispersion $\sigma=|\textbf{v}_\mathrm{circ}|/\sqrt{2}$. For the dark disk, the velocity lag $\textbf{v}_{lag} =(0, 50, 0)$\, km s$^{-1}$ replaces $\textbf{v}_\mathrm{circ}$ and a dispersion of $50$\, km s$^{-1}$ is adopted.

\begin{figure}
 \includegraphics[width=0.47\textwidth,height=55mm]{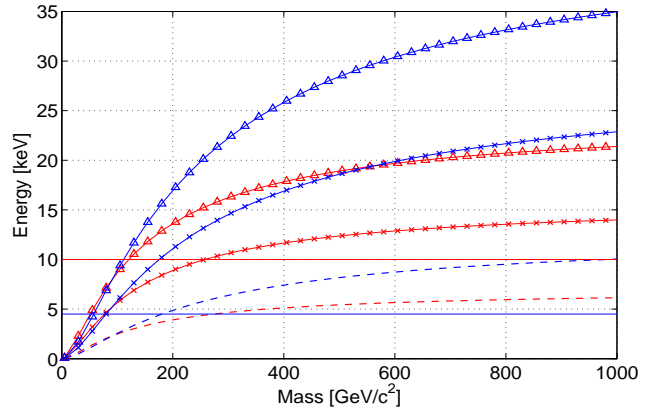}
\caption{Recoil energy below which the signal is dominated by the dark disk (compared with the SHM) as a function of $\MWIMP$ for  Ge (red/gray) and Xe (blue/black) targets. Three different values of $\rhodrat$ (0.5 dashed, 1 {\large{$\times$}} and 2 $\bigtriangleup$) are shown. The horizontal lines mark current experiment thresholds: XENON10 (blue/black) using a Xe and CDMS-II (red/gray) using a Ge target.}
\end{figure}

\begin{figure*}
\begin{center}
 \includegraphics[width=0.33\textwidth]{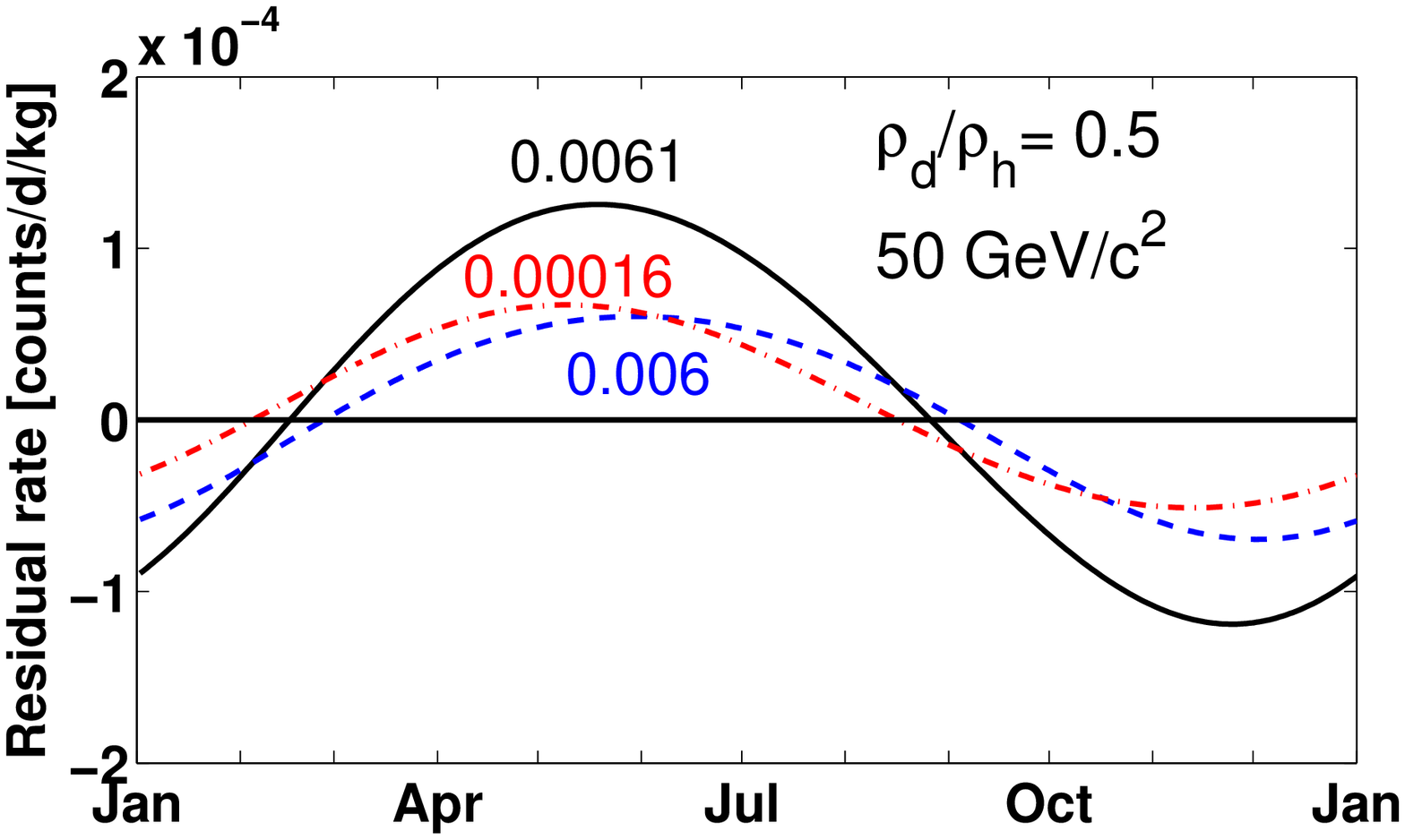} 
 \includegraphics[width=0.33\textwidth,height=39mm]{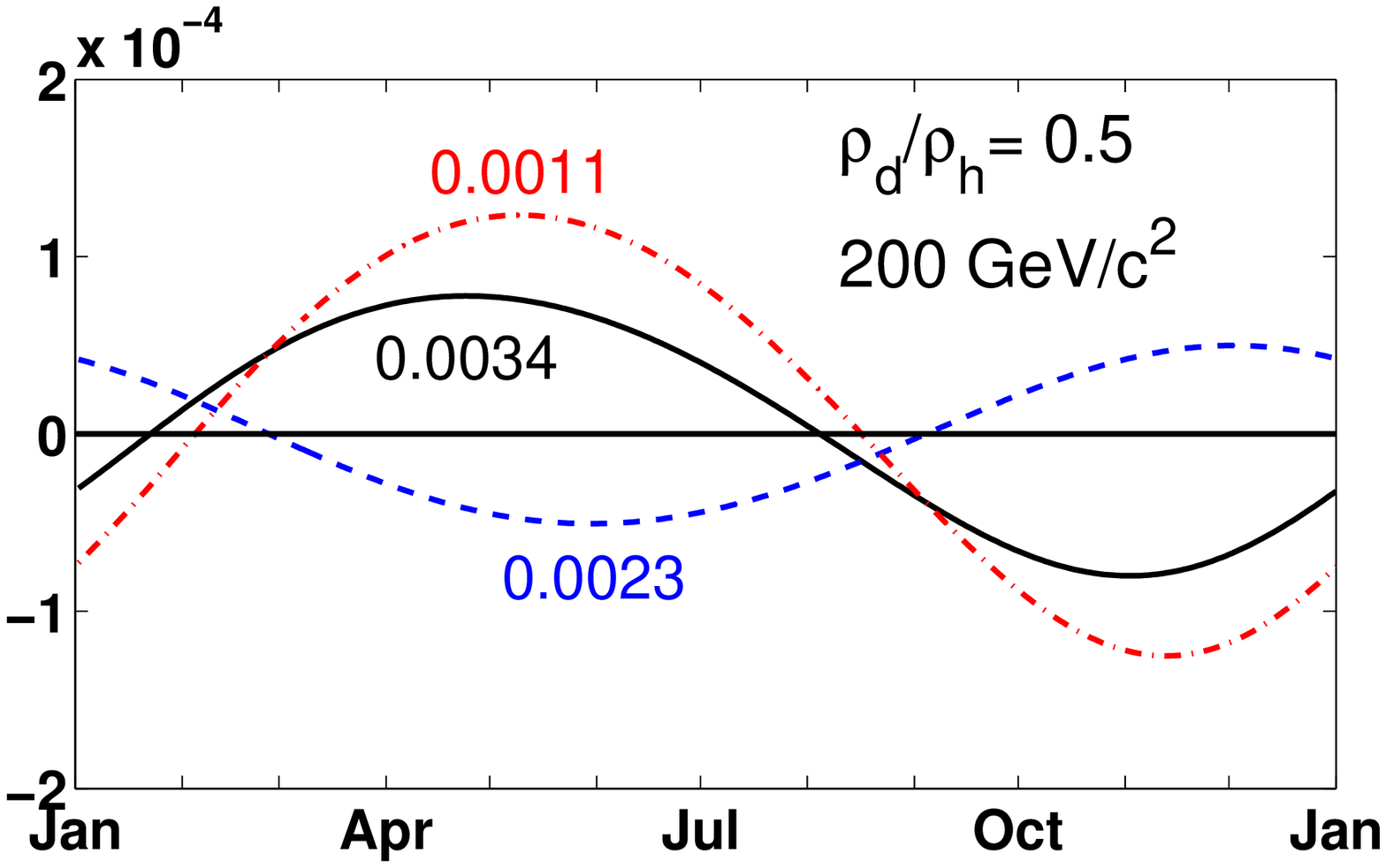} 
 \includegraphics[width=0.33\textwidth,height=39mm]{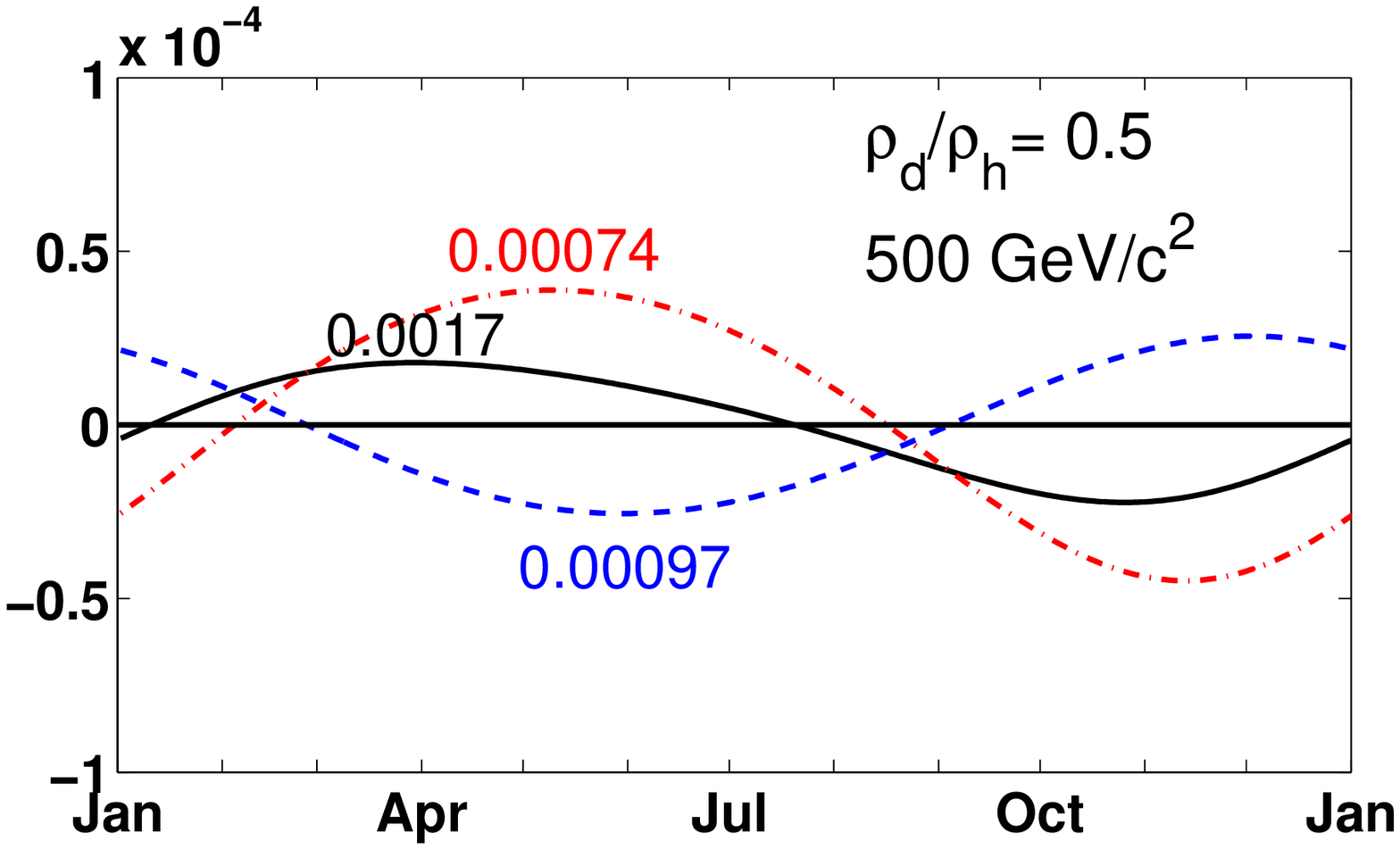} \\
 \includegraphics[width=0.33\textwidth]{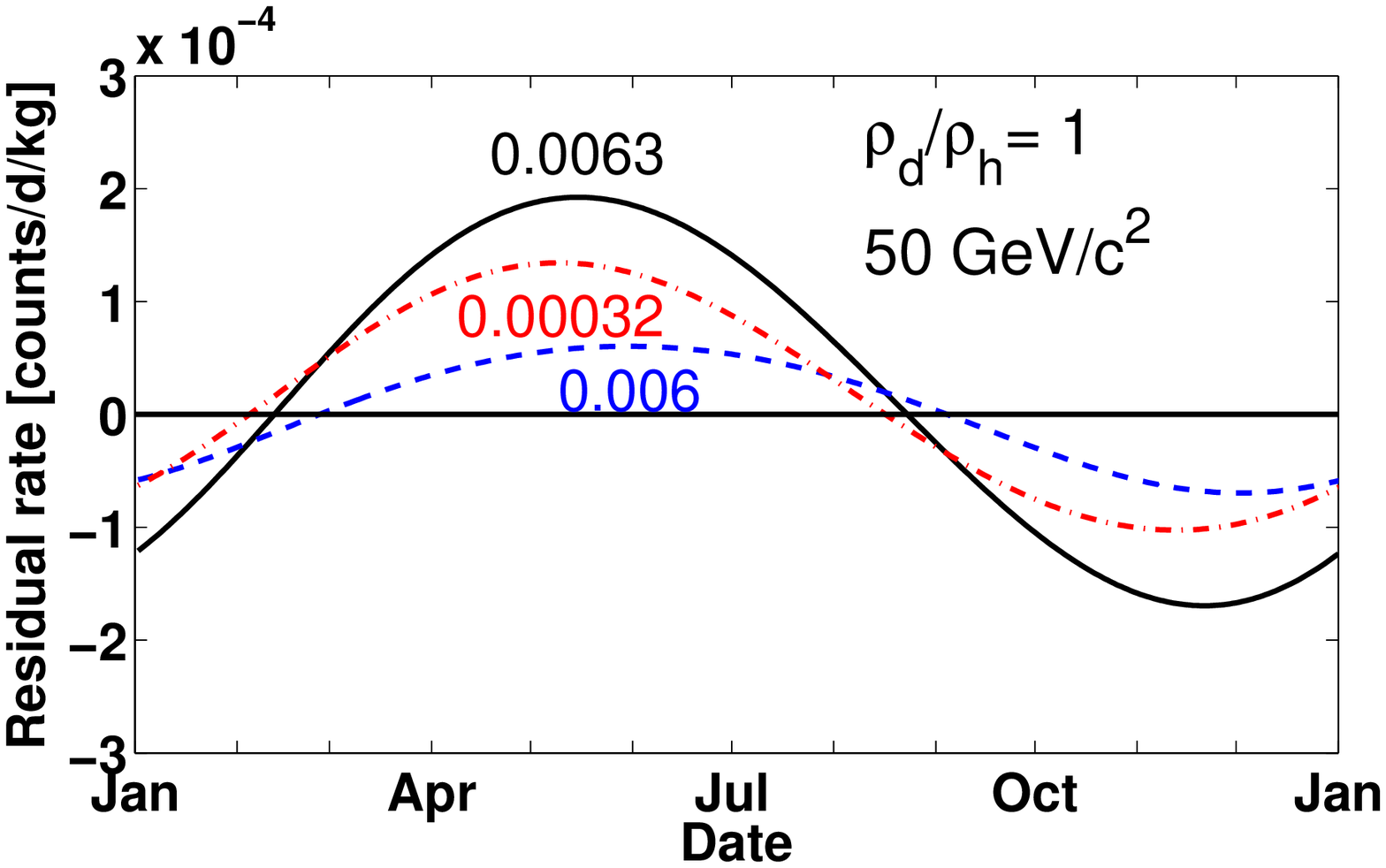} 
 \includegraphics[width=0.33\textwidth,height=39mm]{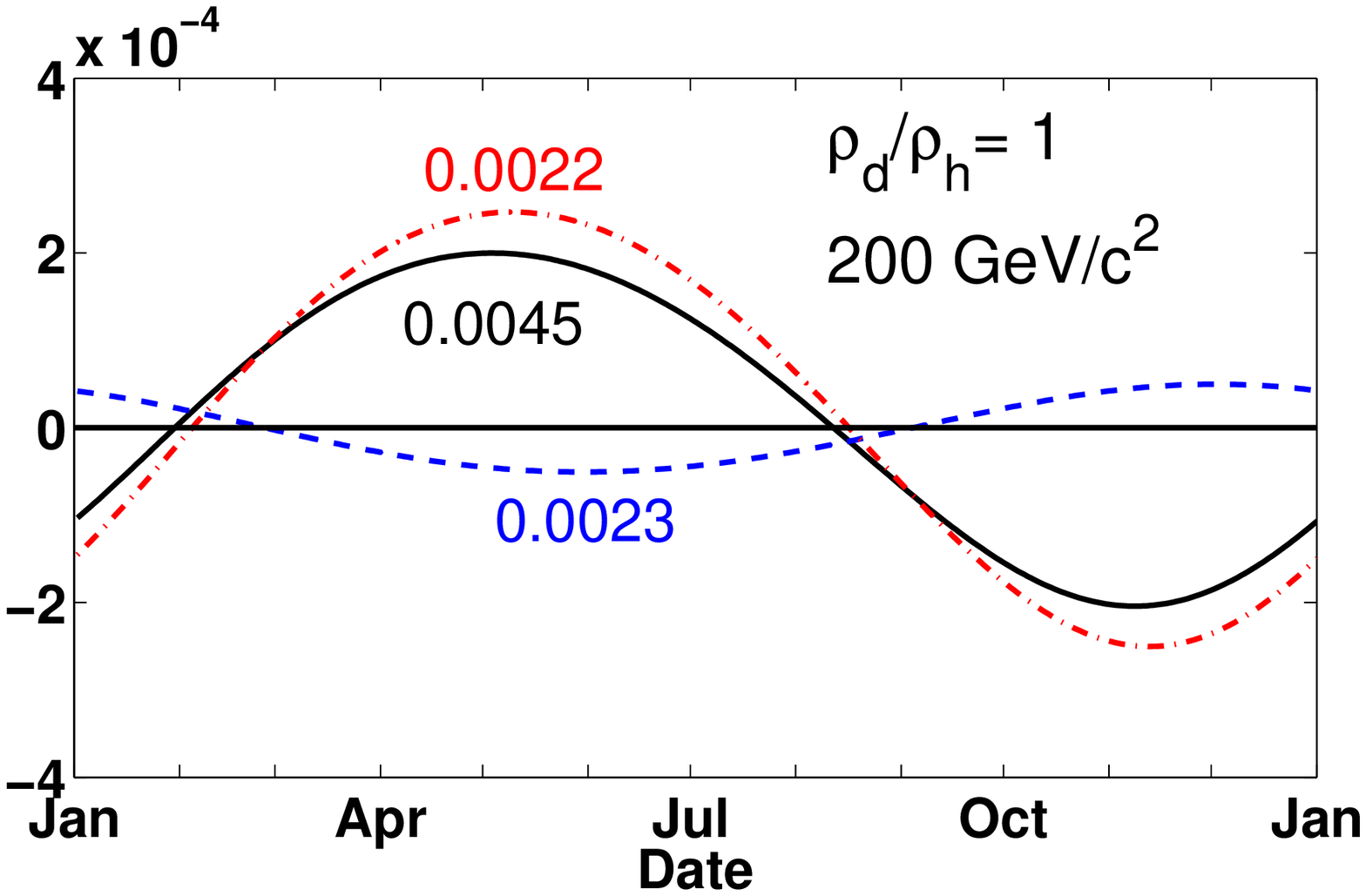} 
 \includegraphics[width=0.33\textwidth,height=39mm]{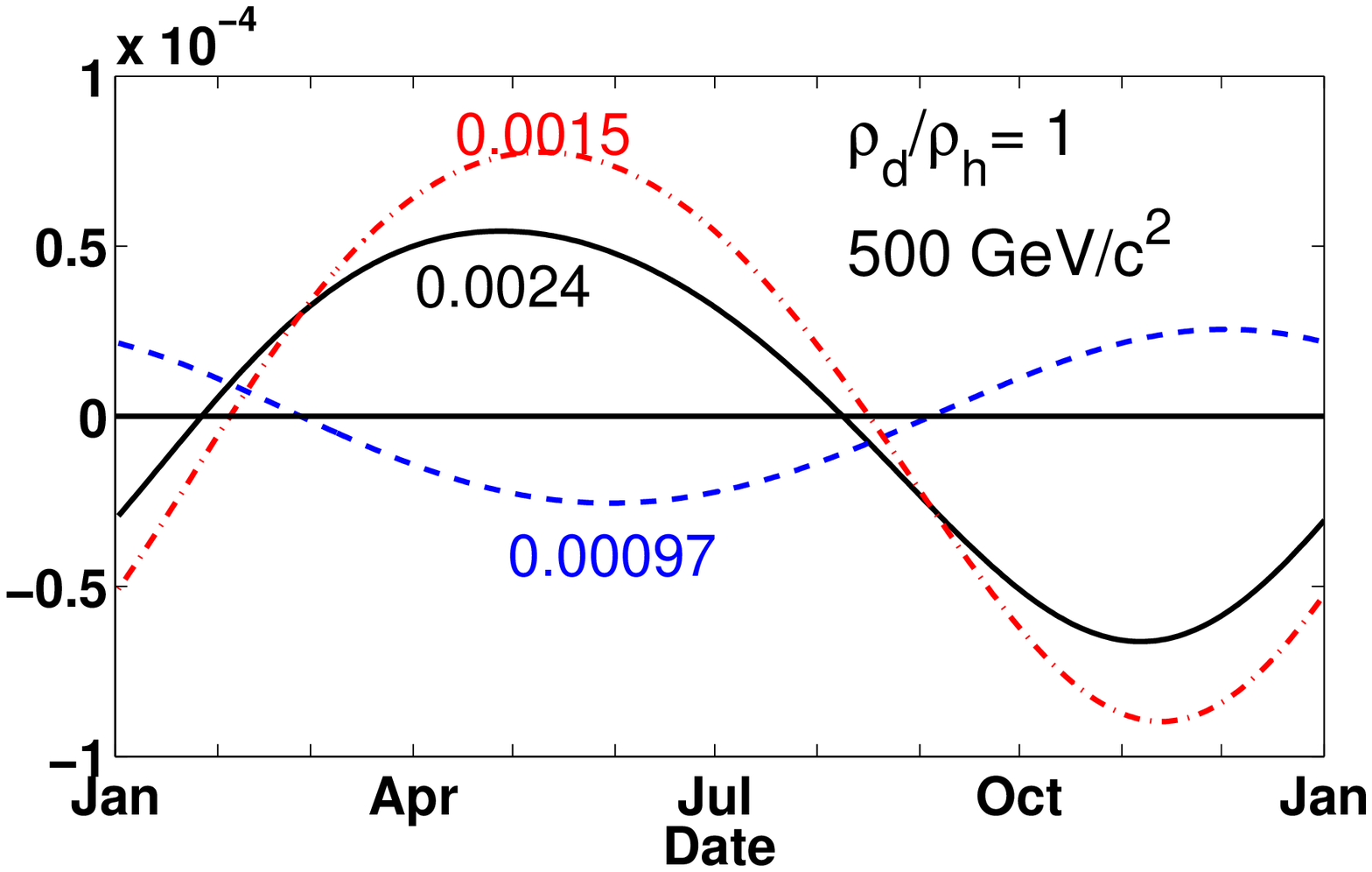} 
\end{center}
\caption{Annual modulation shown as the residual counting rate vs. date for the XENON10 experiment (4.5 to 27 keV). The residuals are calculated with respect to the mean counting rates (given as numbers over each line) using a WIMP-nucleon cross section of 10$^{-8}$ pb. The top/bottom row is calculated for $\rhodrat = 0.5/1$ and $M_{WIMP}$ (left to right) of 50\,\GeV, 200\,\GeV\ and 500\,\GeV. The (blue/dashed) line is the modulation signal obtained from the SHM, the (red/dot-dashed) line is the modulation signal from the dark disk and the (black/solid) line is the total modulation signal. The maximum of the dark disk contribution is shifted to May 9th compared with the SHMs maximum/minimum on May 30th. \textit{Note the different vertical scales in each of the three columns of the plot array.}}
\end{figure*}

\begin{figure}
 \includegraphics[width=0.47\textwidth]{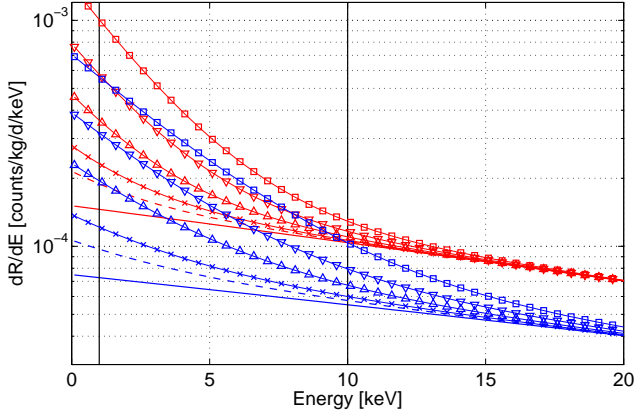}
\caption{Total differential rate for a Ge target in the SHM (solid) and four different  values of $\rhodrat$ (0.1 dashed, 0.2 {\large{$\times$}}, 0.5 $\bigtriangleup$, 1 $\bigtriangledown$ and 2 $\square$) are shown for two $\MWIMP$ (100\,\GeV \,  (red/gray) and 200\,\GeV \,(blue/black)). The vertical lines mark the current CDMS-II threshold and a threshold of 1 keV.}
\end{figure}

The lower relative velocities of the dark disk significantly increase the differential rate at low energies compared with the SHM rate (Figure 1).  Detection of the dark disk crucially depends on the detector's low energy threshold. The differential rate for a specific WIMP target depends on $\MWIMP$. In Figure 2, we show the energy below which the dark disk dominates the rate as a function of $\MWIMP$, for three values of $\rhodrat$. The total rate in a detector is the sum of the two contributions from the SHM and the dark disk, which dominate at high and low energies, respectively. For $\MWIMP \gtrsim 50$\,\GeV, the dark disk contribution lies above current detector thresholds, giving a much greater change in detection rate with recoil energy compared with the SHM alone.

The total rate in a detector using a Ge target is shown in Figure 4 varying $\rhodrat$ and $\MWIMP$. If the detectors threshold is sufficiently low even an extremely conservative dark disk with $\rhodrat=0.1$ can be detected. Current germanium detectors achieve thresholds below 1 keV \citep{Texono,CoGeNT}. The details of the differential rate with energy, as shown in Figure 4, betray both the contribution of the dark disk relative to the SHM and $\MWIMP$. This introduces a mass-dependend characteristic shape of the differential rate which will improve the constrains on $\MWIMP$ upon detection.

The motion of the Earth around the Sun gives rise to an annual modulation of the event rate and recoil energy spectrum \citep{DrukierFreese}. The annual modulation is more pronounced for the dark disk, since the relative change to the mean streaming velocity owing to the Earth's motion is larger ($\sim$19\%) compared with the SHM ($\sim$6\%). We show in Figure 3 the residual integrated rates for a liquid xenon detector throughout a year, for three different $\MWIMP$ and two values of $\rhodrat$. The residuals are calculated with respect to the mean counting rates in a given energy region.

\begin{figure}
 \includegraphics[width=0.47\textwidth,height=55mm]{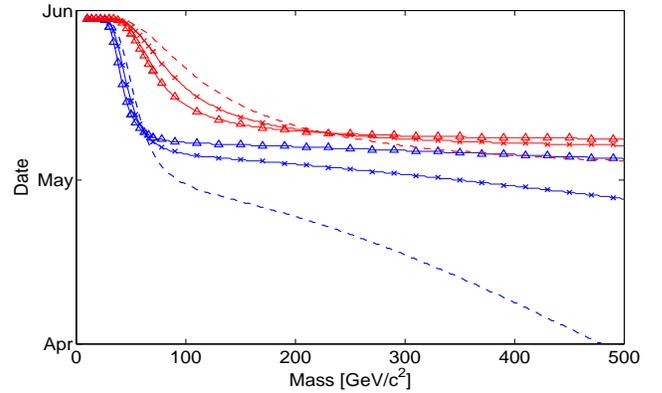}
\caption{Phase shifts as a function of $\MWIMP$ in the energy range reported by the CDMS-II experiment (10-100 keV; red/grey) and the XENON10 experiment (4.5-27 keV; blue/black), for three different values of $\rhodrat$ (0.5 dashed, 1 {\large{$\times$}} and 2 $\bigtriangleup$).}
\end{figure} 

The phase (defined at maximum rate) of the dark disk and the SHM differ because the Sun's motion is slightly misaligned to the dark disk. While the phase of each component does not depend on $\MWIMP$, their sum does because their amplitudes depend on $\MWIMP$. We show this dependency in Figure 5, for three values of $\rhodrat$. The phase shift is determined by the relative contributions of each component. Figure 3 shows that the phase shift is largest for low $\rhodrat$, since in this case the sum preferentially follows the halo modulation phase, while for higher $\rhodrat$ the disk component dominates the modulation phase. This is a new effect introduced by the presence of the dark disk that allows $\MWIMP$ to be 
uniquely determined from the phase of the modulation signal, for given $\rhodrat$. 
Note that there is an amplitude flip for the SHM that occurs as $\MWIMP$ is increased, which is not seen for the dark disk. As $\MWIMP$ is lowered, the ``crossing energy" at which the differential rates for minimal
and maximal WIMP velocity are equal shifts to lower energies. For the dark disk, it remains close to, or below, current thresholds and so the amplitude flip is not seen.

The effect of the dark disk on current upper limits on the SI WIMP-nucleon cross section is shown in Figure 6 for CDMS-II and XENON10 \citep{CDMS, XENON10}. Depending on $\rhodrat$, we exclude new regions in the allowed parameter space for $\MWIMP \gtrsim 50$\,\GeV.

On a final note, we find that including the dark disk component does not change the interpretation of the annual modulation signal observed in the DAMA \citep{DAMA} experiment for pure SI coupling. At high $\MWIMP$, the allowed DAMA region is still excluded by CDMS-II and XENON10 \citep{CDMS, XENON10} results, while at low $\MWIMP$ no new parameter region opens.

\begin{figure}
 \includegraphics[scale=0.44]{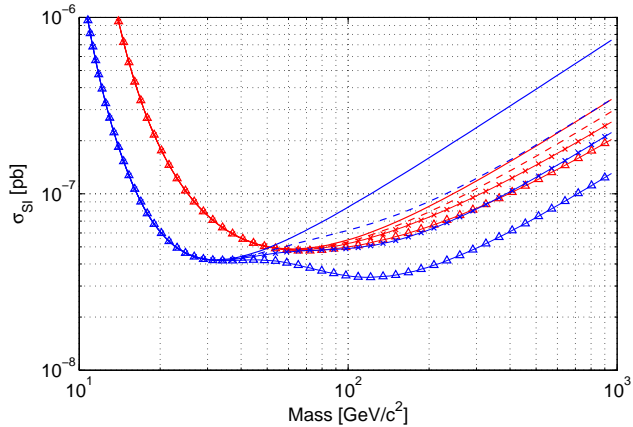}
\caption{Effect of the increased dark matter flux on the SI WIMP-nucleon cross section constraints obtained by the CDMS-II (red/gray) and XENON10 (blue/black) experiments, for three different values of $\rhodrat$ (0.5 dashed, 1 {\large{$\times$}} and 2 $\bigtriangleup$). The solid lines give the constraints if only the SHM component is considered.}
\end{figure}

\section{Conclusions}

In \lcdm, a dark matter disk forms from the accretion of satellites.  We have shown how its low velocity with respect to the Earth alters the expected rate and annual modulation signal in dark matter detectors. Our main findings are as follows. 

The dark disk boosts the detection rates at low recoil energy. For $\MWIMP \gtrsim 50$\,\GeV, recoil energies of 5 - 20 keV and
$\rhodrat \leq 1$, the rate is boosted by factors up to 2.4 for Ge and 3 for Xe targets.  Comparing this with the rates at higher energy will constrain $\MWIMP$, particularly for $\MWIMP > 100$\,\GeV.
 
The dark disk has a different annual modulation phase than the dark halo, while the relative amplitude of the two components varies with recoil energy and $\MWIMP$. As a result, there is a new richness in the annual modulation signal that varies uniquely with $\MWIMP$, for given dark disk properties (the properties of the dark disk will be measured from next generation surveys \citep{Rave06, GAIA}).

The increased expected dark matter flux provides new constraints on the WIMP cross section from current experiments. For likely dark disk properties ($\rhodrat \leq 1$), the constraints for pure SI coupling improve by up to a factor of 1.4 for CDMS-II \citep{CDMS} and 3.5 for XENON10 \citep{XENON10}. 

\acknowledgments

We acknowledge support from the Swiss NSF and the wonderful working environment and support
of UZH.

\end{document}